\documentclass[a4paper]{jpconf}
\usepackage{graphicx}
\usepackage{bm}

\newcommand{\eng}{ErNi$_{2}$Ge$_{2}$}
\newcommand{\tbnige}{TbNi$_{2}$Ge$_{2}$}
\newcommand{\tcs}{ThCr$_{2}$Si$_{2}$}
\newcommand{\rtx}{RT$_{2}$X$_{2}$}
\newcommand{\prcosi}{PrCo$_{2}$Si$_{2}$}

\newcommand{\tbrusi}{TbRu$_{2}$Si$_{2}$}

\newcommand{\TN}{T_{\mbox{\scriptsize N}}}

\begin{document}
\title{Anisotropic magnetic diffuse scattering in an easy-plane type antiferromagnet \eng}

\author{Y Tabata$^{1}$, T Yamazaki$^{1}$, M Okue$^{1}$, H Nakamura$^{1}$ and 
M Matsuura$^{2,}$\footnote[3]{Present address: Graduate School of Science, Osaka University, Toyonaka, Osaka 560-0043, Japan}}

\address{$^{1}$Department of Materials Science and Engineering, Kyoto University, Kyoto 606-8501, Japan}
\address{ $^{2}$Institute for Solid State Physics, University of Tokyo, Kashiwa 277-8581, Japan}

\ead{y.tabata@ht4.ecs.kyoto-u.ac.jp}

\begin{abstract}
We report on neutron scattering studies of a rare earth intermetallic compound \eng . Polarized neutron scattering experiments revealed that the magnetic ordered moment ${\bm m}$ lies in ab-plane. Taking account of a lack of the third higher harmonic reflection, \eng\ is considered to have a helical magnetic structure. The magnetic scattering profiles along the $[100]^{\ast}$- and the $[110]^{\ast}$-directions are well described by the sum of Gaussian and modified-Lorentzian terms, even far below $T_{\mbox{\scriptsize N}}$, indicating that short-range orders coexist with a long-range order. Interestingly, the modified-Lorentzian-type diffuse scattering is not present in the profiles along the $[001]^{\ast}$-direction. The anisotropy of the diffuse scattering suggests that the short-range-order consists of one dimensional long-range helices along the c-axis. 
\end{abstract}

Long-range interactions in magnetic systems, such as the dipolar interaction and the RKKY interaction, cause diverse magnetism because of their competing nature. For instance, Ising spin systems, such as \prcosi\ \cite{prco2si2} and CeSb \cite{cesb}, reveal long-period magnetic soliton structures and complicated successive phase transitions in temperature ($T$) and magnetic field ($H$). On the other hand, long-period helically modulated magnetic structures are realized in Heisenberg and XY spin systems, such as Ho \cite{Ho} and MnAu$_{2}$ \cite{mnau2}.

Rare-earth intermetallic compounds \rtx\ with the tetragonal \tcs -structure (space group $I4/mmm$) are good examples for studying the competing nature of  the long-range RKKY interaction in the Ising-type, the XY-type and the Heisenberg-type spin systems. In the \rtx -system, the magnetic anisotropy varies from the Ising-type to the XY-type systematically as magnetic rare-earth R$^{3+}$-ions being heavier \cite{budko_erni2ge2}. In spite of the simple crystallographic structure, the \rtx -system exhibits rich magnetic properties, such as incommensurate magnetic structures and complicated $HT$ phase diagrams \cite{prco2si2,tbni2ge2,dyru2si2}. 

\eng\ is a candidate for the XY-type spin system in the \rtx -system. Magnetization measurements by using a single crystalline sample revealed easy-plane type magnetic anisotropy and an antiferromagnetic phase transition at $\TN$ $=$ 2.6 K \cite{budko_erni2ge2}. The tetragonal ab-plane is the magnetic easy plane and the magnetization along the c-axis is less than 1/10 of the ab-plane magnetization. According to neutron powder diffraction experiments \cite{andre_erni2ge2}, \eng\ has a long-period sinusoidally modulated magnetic structure with the magnetic wave vector ${\bm k}_{\scriptsize mag}$ $\simeq$ $(0,0,0.75)$ and the magnetic ordered moment forms an angle 64$^{\circ}$ with the c-axis.  This magnetic structure is inconsistent with strong easy-plane type anisotropy observed in the magnetization measurements. In ref.\cite{andre_erni2ge2}, the helical structure also arises as a possible magnetic structure. 

In this article, we report on our recent results of polarized and unpolarized neutron scattering studies by using single crystalline samples of \eng . Polarized neutron scattering experiments exhibit that the magnetic ordered moment in \eng\ lies in the ab-plane.  Successive phase transitions do not arise in contrast with the case of Ising-type systems as \tbnige\ \cite{tbni2ge2}. Instead, a novel anisotropic magnetic diffuse scattering was found in the antiferromagnetic ordered phase. 

Single crystalline samples of \eng\ were grown by the Czochralski method with a tetra-arc furnace. Magnetization measurements were performed by using a SQUID magnetometer MPMS (Quantum Design). Polarized and unpolarized neutron scattering experiments were performed on the triple-axis spectrometer PONTA and HQR at the JRR-3M reactor in JAEA, respectively. The polarized neutron energy $E$ $=$ 13.5 meV was chosen by Heusler monochromator and analyzer. The direction of the polarization vector of neutron spins ${\bm P}$ was tuned to be parallel or perpendicular to the scattering vector ${\bm K}$ by using Helmholtz coils. The polarization ratio of the neutron spins was estimated from the nuclear Bragg peak at $(0,0,4)$ as being $\sim$ $20$. The unpolarized neutron energy $E$ $=$ 13.5 meV was selected by PG monochromator and analyzer. A PG filter was used to eliminate higher harmonic reflections from the monochromator. 

\begin{figure}[h]
\includegraphics[width=24pc]{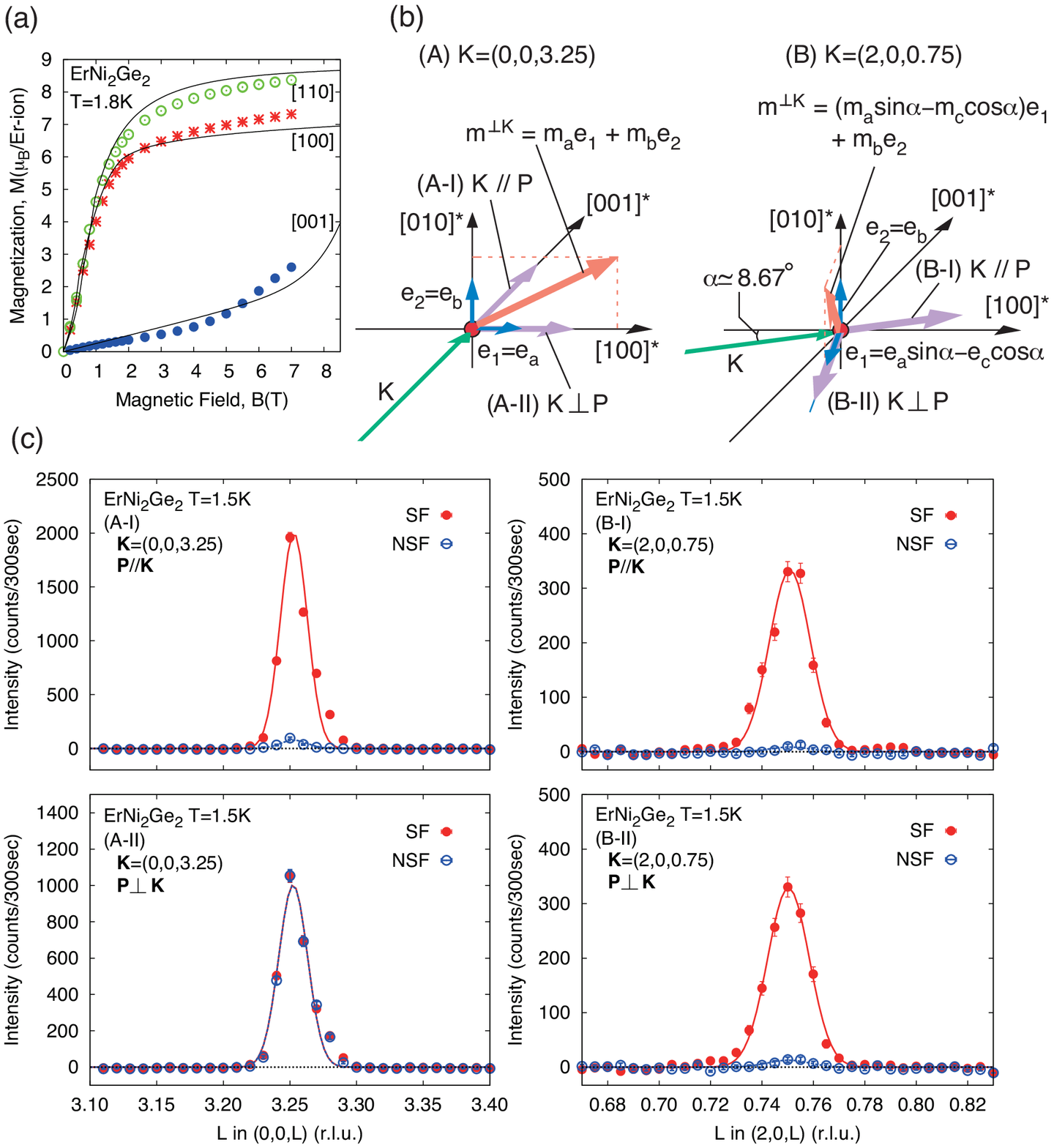}\hspace{1pc}%
\begin{minipage}[b]{13pc}
\caption{\label{fig1} (a) Magnetization of single crystalline \eng\  at 1.8 K. Solid lines represent calculated results based on the 2-sublattice MF model with the CEF. (b) The configurations of the ${\bm K}$, the ${\bm P}$ and the ${\bm m}^{\perp {\bm K}}$ in the present experimental conditions. ${\bm e}_{1}$ and ${\bm e}_{2}$ are the unit vectors perpendicular to ${\bm K}$ in and perpendicular to the scattering plane, respectively. The scattering  plane is the horizontal (H0L)-plane. $m_{\mbox{\scriptsize a}}$, $m_{\mbox{\scriptsize b}}$ and $m_{\mbox{\scriptsize c}}$ represent the components of ${\bm m}$ along the a-, b- and c-axes, respectively. (c) L-scan profiles of the spin-flip (SF) and the non-spin-flip (NSF) scattering by using polarized neutron.  (A-I) ${\bm K}$ $=$ $(0,0,3.25)$, ${\bm P} \parallel {\bm K}$.  (A-II) ${\bm K}$ $=$ $(0,0,3.25)$, ${\bm P} \perp {\bm K}$.  (B-I) ${\bm K}$ $=$ $(2,0,0.75)$, ${\bm P} \parallel {\bm K}$.  (B-II) ${\bm K}$ $=$ $(2,0,0.75)$, ${\bm P} \perp {\bm K}$.}
\end{minipage}
\end{figure}

\begin{table}[h]
\caption{\label{tab1} List of the components of the ${\bm m}^{\perp {\bm K}}$ perpendicular and parallel to ${\bm P}$, ${\bm m}^{\perp {\bm K}}_{\perp {\bm P}}$ and ${\bm m}^{\perp {\bm K}}_{\parallel {\bm P}}$ in the experimental conditions illustrated in Fig. \ref{fig1}(b).}
\begin{tabular}{ccccc}
\br
 & (A-I) &  (A-II) &  (B-I) &  (B-II) \\ \mr
${\bm m}^{\perp {\bm K}}_{\perp {\bm P}}$ & $m_{\mbox{\scriptsize a}} {\bm e}_{1}  + m_{\mbox{\scriptsize b}} {\bm e}_{2}$ & $m_{\mbox{\scriptsize b}} {\bm e}_{2}$ & $(m_{\mbox{\scriptsize a}} \sin \alpha - m_{\mbox{\scriptsize c}} \cos \alpha ) {\bm e}_{1} + m_{\mbox{\scriptsize b}} {\bm e}_{2}$ & $m_{\mbox{\scriptsize b}} {\bm e}_{2}$ \\ \mr
${\bm m}^{\perp {\bm K}}_{\parallel {\bm P}}$ & $-$ & $m_{\mbox{\scriptsize a}} {\bm e}_{1}$ & $-$ & $(m_{\mbox{\scriptsize a}} \sin \alpha - m_{\mbox{\scriptsize c}} \cos \alpha ) {\bm e}_{1}$ \\
\br
\end{tabular}
\end{table}

Magnetization processes along the $[100]$-, the $[110]$- and the $[001]$-axes at 1.8 K are shown in Fig. \ref{fig1} (a). As reported in ref. \cite{budko_erni2ge2}, strong easy-plane type magnetic anisotropy was found. In magnetic easy ab-plane, weak magnetic anisotropy exists and the $[110]$-axis is the magnetic easy axis. Similar in-plane anisotropy was found in several Er-compounds in the \rtx -system \cite{erru2si2,erru2ge2}.  Solid lines  in Fig.\ref{fig1} (a) are calculated results based on the 2-sublattice mean-field (MF) Hamiltonian with the crystalline-electric-field (CEF) $\mathcal{H}_{\mbox{\scriptsize MF}} + \mathcal{H}_{\mbox{\scriptsize CEF}}$. The magnetic anisotropy of \eng\ is well described by the CEF effect in the tetragonal symmetry. The N\'{e}el temperature $\TN$ was estimated to be 3.0 K, being slightly higher than that reported previously \cite{budko_erni2ge2}.

Figure \ref{fig1} (c) shows L-scan profiles of spin-flip (SF) and non-spin-flip (NSF) scatterings at 1.5 K in the polarized neutron scattering experiments around the scattering vector (A) ${\bm K}$ $=$ $(0,0,3.25)$ ($=$ ${\bm G}_{004} - {\bm k}_{\mbox{\scriptsize mag}}$) and (B) ${\bm K}$ $=$ $(2,0,0.75)$ ($=$ ${\bm G}_{200} + {\bm k}_{\mbox{\scriptsize mag}}$) in the condition of the ${\bm P}$ as (I) ${\bm P} \parallel {\bm K}$ and (II) ${\bm P} \perp {\bm K}$. Only the component of magnetic moment ${\bm m}$ perpendicular to ${\bm K}$, ${\bm m}^{\perp {\bm K}}$, contributes to the magnetic neutron scattering. In the polarized neutron scattering experiments, the component of ${\bm m}^{\perp {\bm K}}$ perpendicular and parallel to ${\bm P}$, $m^{\perp {\bm K}}_{\perp {\bm P}}$ and $m^{\perp {\bm K}}_{\parallel {\bm P}}$, contribute to the SF and the NSF scattering respectively in a simple manner. The configurations of the ${\bm K}$, the ${\bm P}$ and the ${\bm m}^{\perp {\bm K}}$ in the present experimental conditions are illustrated in Fig. \ref{fig1} (b), and the $m^{\perp {\bm K}}_{\perp {\bm P}}$ and the $m^{\perp {\bm K}}_{\parallel {\bm P}}$ are listed in table \ref{tab1}. In the (A-I)- and the (B-I)-conditions, only the SF scattering channel was observed, because the scatterings at $(0,0,4)$ and $(2,0,0.75)$ are purely magnetic. Observed very weak NSF scatterings are due to the incompleteness of the neutron polarization. The intensities of the SF and the NSF scattering are almost the same in the (A-II)-condition, indicating the isotropic magnetic ordered moment in ab-plane. The NSF scattering in the (B-II)-condition was not observed any more than that in the (A-I)- or the (B-I)-condition was. Because the lattice constant $c$ is much longer than $a$, $c$ $=$ $9.769$ \AA\ and  $a$ $=$ $4.019$ \AA , the scattering vector in the (B)-condition is almost parallel to the $[100]^{\ast}$-axis, i.e. $\sin \alpha$ $=$ $0.152$ and $\cos \alpha$ $=$ $0.988$ where $\alpha$ is an angle between the $[100]^{\ast}$-axis and the scattering vector. Hence, the NSF scattering  in the (B-II) condition is contributed by the c-component of the magnetic ordered moment $m_{\mbox{\scriptsize c}}$ mostly and the lack of the NSF scattering indicates the lack of $m_{\mbox{\scriptsize c}}$. $m_{\mbox{\scriptsize c}}$ was estimated to be 1/10 of the component in the ab-plane $m_{\mbox{\scriptsize ab}}$ at most, being consistent with the anisotropy observed in the single crystal magnetization measurements. Magnetic scattering at the 3rd higher harmonic position, ${\bm K}$ $=$ ${\bm G}_{004} - 3{\bm k}_{\mbox{\scriptsize mag}}$ $=$ $(0,0,1.75)$, was hardly observed, whose intensity is less than $1/1000$ of that at $(0,0,3.25)$. Hence, the possible magnetic modulation is either the purely sinusoidal or the helical one. Taking account of localized nature of magnetic f-electrons of Er-ions, the helical modulation is most probable. 

\begin{figure}[h]
\includegraphics[width=36pc]{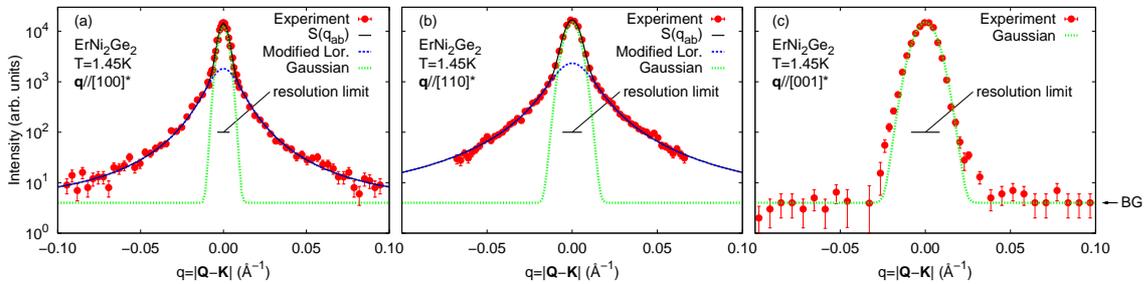}%
\caption{\label{fig2} Scan profiles of the magnetic scattering with the scattering vector ${\bm Q}$ $=$ ${\bm K} + {\bm q}$ and ${\bm K}$ $=$ $(0, 0, 3.25)$, where (a) ${\bm q} \parallel [100]^{\ast}$. (b) ${\bm q} \parallel [110]^{\ast}$.  ${\bm q} \parallel [001]^{\ast}$. The functional form of $S(q_{\mbox{\scriptsize ab}})$ is described in Eq. (\ref{eq1}) in the text.}
\end{figure}

Figure \ref{fig2} shows magnetic scattering profiles at $T$ $=$ $1.45$ K with the scattering vector ${\bm Q}$ $=$ ${\bm K} + {\bm q}$ where ${\bm K}$ $=$ $(0,0,3.25)$ in the unpolarized neutron scattering experiments. The magnetic scattering extends widely along the $[100]^{\ast}$- and the $[110]^{\ast}$-directions, as shown in Figs. \ref{fig2} (a) and (b).  On the other hand, the profile along the $[001]^{\ast}$-direction shown in Fig. \ref{fig2} (c) exhibits a narrow peak, being considered in the resolution limit. The results indicate existence of an anisotropic short-range order whose correlation length in the ab-plane $\xi _{\mbox{\scriptsize ab}}$ is finite and that along the c-axis $\xi _{\mbox{\scriptsize c}}$ is infinite or too long to be observed in the present experimental conditions. For a numerical analysis of the scattering profiles along the $[100]^{\ast}$- and the $[110]^{\ast}$-axes, we used the following functional form $S(q_{\mbox{\scriptsize ab}})$ with the Gaussian resolution functions, 
\begin{equation}
 \label{eq1}
 S(q_{\mbox{\scriptsize ab}}) = S_{0}^{2} \delta (q_{\mbox{\scriptsize ab}}) + \frac{D_{0}}{\left[ 1+ (q_{\mbox{\scriptsize ab}}/\kappa_{\mbox{\scriptsize ab}})^{2} \right]^{\theta}}  \, ,
\end{equation}
where $\kappa_{\mbox{\scriptsize ab}}$ ($=$ $2\pi/\xi_{\mbox{\scriptsize ab}}$) and $q_{\mbox{\scriptsize ab}}$ are an inverse correlation length and a component of ${\bm q}$ in the ab-plane, respectively. The first term represents the magnetic Bragg reflection at ${\bm K}$, giving a Gaussian component with a width of the resolution limit. The second term is a modified Lorentzian, being a simple standard Lorentzian when the exponent $\theta$ $=$ 1, which represents the anisotropic diffuse component. This form describes well the experimental data as shown in Figs. \ref{fig2} (a) and (b). From the analysis with Eq. (\ref{eq1}), $S_{0}^{2}$ $=$ $53.75 \pm 0.90$, $D_{0}$ $=$ $(2.00 \pm 0.12) \times 10^{3}$, $\kappa _{\mbox{\scriptsize ab}}$ $=$ $(9.05 \pm 0.61) \times 10^{-3} \mbox{\AA}^{-1}$ $=$ $(5.78 \pm 0.39) \times 10^{-3} a^{\ast}$ and $\theta$ $=$ $1.28 \pm 0.04$ were obtained. The profiles along the $[001]^{\ast}$-direction can be fitted to only the Gaussian term with a width of the resolution limit, as shown in Fig. \ref{fig2} (c). 

It is an open question what is the origin of the anisotropic magnetic short-range order in \eng . The diffuse scattering component was not observed at all above and below $\TN$ in nuclear scattering profiles. Hence, the observed magnetic short-range order does not come from neither crystallographic disorder nor chemical distortion driven by the antiferromagnetic transition through magnetoelastic couplings. The anisotropy of the diffuse scattering strongly suggests that the magnetic short-range order consists of one dimensional long-range helices along the c-axis, and the `phase' of the helix fluctuates with ferromagnetic correlations in the ab-plane. It looks similar to the Kosterlitz-Thouless phase in two dimensional ferromagnetic XY spin systems \cite{KT}. The frustration coming from the competing long-range interactions in the ab-plane could derive such short-range order. In contrast, the Ising system, i.e. \tbrusi\ \cite{tbru2si2}, reveals quite complicated, but being long-ranged, two dimensional patterns in the ab-plane. In \eng , the XY-like continuous freedom of the phase of the helix may allow existence of the short-range order with the finite correlation length $\xi_{\mbox{\scriptsize ab}}$ $\sim$ $170a$ in the long-range order. 

%

\end{document}